\pdfoutput=1
\documentclass[11pt]{article}

\usepackage{graphicx}
\usepackage{amsmath}
\usepackage{amssymb}
\usepackage{dcolumn}
\usepackage{bm}
\usepackage{slashed}
\usepackage{color}
\usepackage{authblk}

\newcommand{\be}{\begin{equation}}
\newcommand{\ee}{\end{equation}}
\newcommand{\ba}{\begin{eqnarray}}
\newcommand{\ea}{\end{eqnarray}}
\newcommand{\no}{\nonumber \\}
\newcommand{\gsim}{\mathrel{\hbox{\rlap{\lower.55ex \hbox {$\sim$}}
                   \kern-.3em \raise.4ex \hbox{$>$}}}}
\newcommand{\lsim}{\mathrel{\hbox{\rlap{\lower.55ex \hbox {$\sim$}}
                   \kern-.3em \raise.4ex \hbox{$<$}}}}

\def\roughly#1{\mathrel{\raise.3ex\hbox{$#1$\kern-.75em%
\lower1ex\hbox{$\sim$}}}}
\def\lsim{\roughly<}
\def\gsim{\roughly>}

\def\({\left(}
\def\){\right)}
\def\[{\left[}
\def\]{\right]}
\def\<{\langle}
\def\>{\rangle}

\def\l{{\lambda}}
\def\L{{\Lambda}}
\def\d{{\delta}}
\def\D{{\Delta}}
\def\o{{\omega}}

\def\e{{\epsilon}}

\def\a{{\alpha}}
\def\b{{\beta}}

\def\g{{\gamma}}

\def\h{\eta}

\def\m{{\mu}}
\def\n{{\nu}}
\def\r{{\rho}}
\def\s{{\sigma}}
\def\S{{\Sigma}}

\def\th{{\theta}}

\def\ps{{\psi}}

\newcommand{\pd}{{\partial}}

\newcommand{\tr}{\text{tr}}

\newcommand{\eq}{\text{eq}}

\date{\today}

\begin{document}

\title{\bf Steady state, displacement current and spin polarization for massless fermion in a shear flow}

\author[1]{Shu Lin \thanks{linshu8@mail.sysu.edu.cn}}
\affil[1]{School of Physics and Astronomy, Sun Yat-Sen University, Zhuhai 519082, China}
\author[2]{Ziyue Wang \thanks{Zi-yue.Wang@outlook.com}}
\affil[2]{School of Physics and Optoelectronic Engineering, Beijing University of Technology, Beijing 100021, China}

\maketitle

\begin{abstract}
We consider spin polarization of massless fermions in a shear flow, whose complete contributions contain magnetization current and side-jump current known from collisional chiral kinetic theory. We argue that the side-jump current adopts interpretation of displacement current. We explicitly determine the displacement current contribution in the steady state reached in shear flow for a QED plasma. We find the displacement contribution enhances the magnetization contribution at small and large momenta, but leads to a suppression effect at intermediate momenta. Major differences from previous studies on collisional effect are: (i) the fermions are in the same steady state as the medium rather than being probes; (ii) Compton scattering and pair annihilation are also included in addition to the Coulomb scattering considered before. 
\end{abstract}


\newpage

\section{Introduction and summary}

It has been realized that large orbital angular momentum in off-central heavy ion collisions leads to global polarization of quark-gluon plasma (QGP) \cite{Liang:2004ph,Liang:2004xn}. Indeed the observation of global polarization of $\L$ hyperon in heavy ion collisions has witnessed QGP as rapid spinning matter \cite{STAR:2017ckg}. Later studies have found that the global polarization data can be well understood with spin-vorticity coupling \cite{Becattini:2013fla, Fang:2016vpj, Li:2017slc, Liu:2019krs, Becattini:2017gcx, Wei:2018zfb, Wu:2019eyi, Fu:2020oxj, Zhang:2019xya, Weickgenannt:2020aaf}. Predictions for local polarization have been made in \cite{Becattini:2017gcx,Wei:2018zfb,Fu:2020oxj}, which turn out to differ from measurement by the sign \cite{STAR:2019erd}. In fact, quantum kinetic theory as well as field theory studies indicate that spin can couple to all other fluid gradients \cite{Hidaka:2017auj,Liu:2021uhn,Becattini:2021suc}. In particular, the shear contribution has been found as another main source to local polarization, which can potentially give the correct sign \cite{Fu:2021pok,Becattini:2021iol,Yi:2021ryh,Fu:2022myl,Wu:2022mkr}. A more recent experimental study has added a new twist to the story \cite{STAR:2023eck}: measurements of local polarization at high transverse momentum show significantly deviation from the prediction based on vorticity and shear contributions combined. It is clear that we are still far from complete understanding of local polarization phenomenon.

In fact, the question of polarization in a shear flow is much complicated by the following fact: the shear flow drives the system into a steady state with nonequilibrium distribution and detailed balance no longer holds. This is in contrast to the vorticity case where the system remains in equilibrium with detailed balance. It is the deviation from equilibrium and the resulting nonvanishing collision that leads to an extra contribution polarization. Indeed, such a contribution for massless fermion is known as side-jump current \cite{Chen:2015gta,Chen:2014cla,Hidaka:2016yjf}, see also \cite{Sheng:2022ssd,Yang:2020hri} for massive generalization. In the massless limit, the composition of current reads
\begin{align}\label{current}
j^\m=p^\m f+S^{\m\n}\pd_\n f+\int_{BCD}C_{ABCD}\bar{\D}^\m,
\end{align}
The second term on the right-hand side (RHS) is a kinematic contribution considered in phenomenological studies where $f$ corresponds a local equilibrium distribution in a shear flow,
with $S^{\m\n}=\l\frac{\e^{\m\n\r\s}p_\r n_\s}{p\cdot n}$ and $\l$ being helicity of the fermion.
The last term on the RHS is the side-jump current with
\begin{align}\label{jump}
\bar{\D}^\m=\l\frac{\e^{\m\n\r\s}p_\n\bar{n}_\r n_\s}{(p\cdot n)(p\cdot\bar{n})}
\end{align}
being the jump from center of mass frame specified by $\bar{n}^\m=(p_A^\m+p_B^\m)/\sqrt{s}$ to an arbitrary frame specified by $n$. Although each term in \eqref{current} depends on $n^\m$, the dependencies cancel out in the total current. Note that the collision kernel for binary collision $C_{ABCD}$ would vanish by detailed balance in local equilibrium state, but does not in a steady state. The expected cancellation of the $n$-dependencies suggests that the side-jump current may be as important as the other contributions. 
Indeed, in \cite{Lin:2022tma}, we perform an explicit calculation of this collision induced contribution for massive probe fermion in a QED medium. We have found the collisional contribution parametrically the same as the kinematic one and moreover, the collisional contribution leads to a suppression of the kinematic contribution.

The results in \cite{Lin:2022tma} come with two limitations: firstly we have assumed the probe fermion to have equilibrium distribution for technical simplicity, which means the probe fermion itself is not in a steady state but only the medium fermions are; secondly for massive fermions, the analog of the first term of RHS of \eqref{current} is dynamical, which needs to be determined by solving an axial kinetic equation \cite{Hattori:2019ahi,Weickgenannt:2019dks,Gao:2019znl}. In this paper, we improve on these two aspects by considering polarization of massless medium fermions. We will argue based on kinetic equation and \eqref{current} that the collisional contribution could lead to enhancement of the kinematic contribution. We will see this is indeed true for fermions with small and large momenta from explicit calculations. On the other hand, the dynamical contribution will be found to vanish identically in the massless case. This is a consequence of the spin-momentum locking in the massless limit that renders the dynamical contribution trivial.


The paper is structured as follows: in Section~\ref{sec_disp}, we interpret the side-jump current as displacement current. This is analogous to the interpretation of the second term of the RHS of \eqref{current} as magnetization current; in Section~\ref{sec_enhance}, we consider the steady state in a shear flow and argue in this case the displacement current could lead to enhancement of the magnetization current; in Section~\ref{sec_qed} we give an explicit example of massless fermion polarization in a QED medium with shear flow; 
Section~\ref{sec_conclude} is devoted to conclusion and an outlook.

\section{Side-jump current as displacement current}\label{sec_disp}

Let us first recall the interpretation of the term $S^{\m\n}\pd_\n f$ \cite{Chen:2014cla}. In the frame $n^\m=(1,0,0,0)$, we have
\begin{align}
S^{ij}\pd_j f=\nabla\times \l\hat{p}f=\nabla\times{\bf M},
\end{align}
with $M$ being the magnetic moment of massless fermions. This leads to the interpretation of magnetization current. In Maxwell theory, the complete contribution to current also includes displacement current
\begin{align}
{\bf j}=\pd_t{\bf P}+\nabla\times{\bf M},
\end{align}
with $P$ being the electric polarization. Unlike the magnetization current, which exists in non-interacting systems, the displacement current requires interaction. This is in line with the collision dependent side-jump current. We will argue that the side-jump current indeed adopts interpretation of displacement current. For simplicity, we consider homogeneous distribution in frame $n$, i.e. $\pd_i f=0$ but $\pd_tf\ne 0$. In this case, we find the corresponding distribution function $f$ being the same as the counterpart ${\bar f}$ in no-jump frame according to Eq (16) of \cite{Chen:2015gta}
\begin{align}
{\bar f}-f=-\bar{\D}\cdot\pd f+\int_{BCD}\frac{{\bar \D}\cdot{\bar n}}{p\cdot{\bar n}}=0.
\end{align}
The first term on the RHS vanishes because $\pd_i f=0$ and $\bar\D$ has only nonvanishing spatial components in frame $n$. The second term vanishes identically. It follows that the kinetic equation takes the simple form as in the no-jump frame
\begin{align}
p\cdot\pd f=\int_{BCD}C_{ABCD}.
\end{align}
The electric polarization comes from displacement of charge. A natural candidate for it would be $P^i={\bar\D}^i p_0f$ with the side-jump identified as the displacement viewed in frame $n$ and $p_0f$ as density. However, since the side-jump depends on momentum of the other particle in the collision, we need to use an averaged quantity
\begin{align}
P^i=\<{\bar\D}^i\> p_0f.
\end{align} 
The detail of the average will be clear soon. Now the displacement current becomes
\begin{align}
&\pd_tP^i=\pd_t\(\<{\bar\D}^i\> p_0f\)=\<{\bar\D}^i\>p_0\pd_tf\nonumber\\
=&\<{\bar\D}^i\>\int_{BCD}C_{ABCD},
\end{align}
where in the second line we have used the kinetic equation. If we choose the average $\<{\bar\D}^i\>=\frac{\int_{BCD}C_{ABCD}{\bar\D}^i}{\int_{BCD}C_{ABCD}}$, we easily confirm the displacement current interpretation of side-jump current:
\begin{align}
\pd_tP^i=\int_{BCD}C_{ABCD}{\bar\D}^i.
\end{align}

The difference between the right-handed and left-handed currents gives the spin polarization. The difference of the conducting current (first term on the RHS of \eqref{current}) corresponds to the spin carried by individual particles as $\int dp_0\d(P^2)p^i(f_R-f_L)=\hat{p}^i(f_R-f_L)/2$. The counterpart of the magnetization and displacement current correspond to gradient contributions needed to restore frame independence.

\section{Steady state and collisional effect in a shear flow}\label{sec_enhance}

Now we consider fermion in a steady state due to a shear flow. We choose the frame vector to be the fluid velocity vector $n^\m=u^\m$. We are interested in the relative sign of magnetization current and displacement current contributions in local rest frame of the fluid where $n^\m=u^\m=(1,0,0,0)$. The magnetization current contribution reads
\begin{align}\label{mag}
&S^i_{\text{magn}}\sim \(S_{\l=1/2}^{ij}-S_{\l=-1/2}^{ij}\)\pd_j f_V\sim -\e^{ijk}p_kp_l\pd_ju_lf_V'\nonumber\\
&\sim\e^{ijk}p_kp_l\s_{jl},
\end{align}
where we have only kept track of the $p_i$ and $\s_{ij}$ dependencies and the sign (note that $f_V'<0$).

\begin{figure}[htbp]
	\begin{center}
		\includegraphics[height=4.2cm,clip]{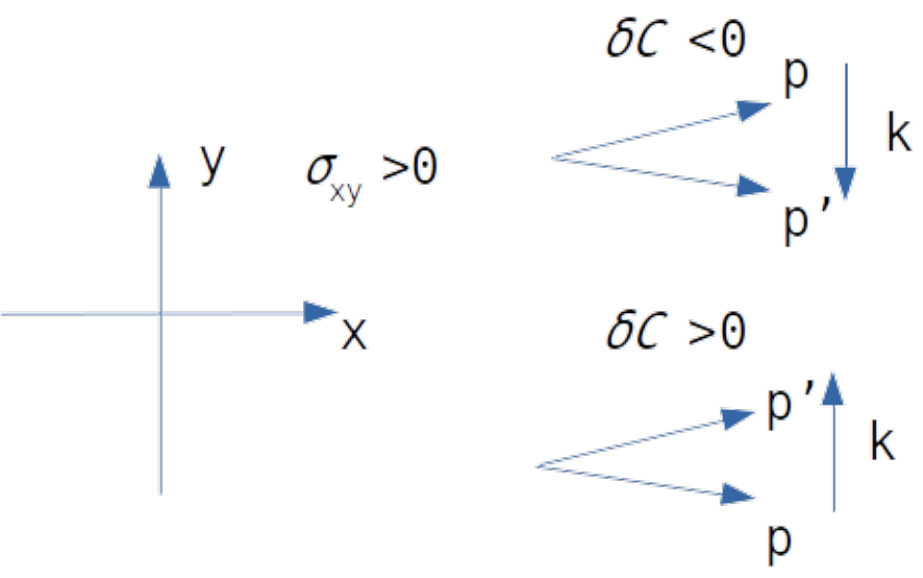}
		\caption{Illustration of average momentum from effective collision in a shear flow with $\s_{xy}>0$. Collision effect on particle with momentum $p$ in the first quadrant and $p_x\gg p_y$ (upper right) is to deflect it into particle with momentum $p'$ in the fourth quadrant through multiple small angle scatterings. The average momentum transfer along $-y$ direction is used as average momentum $k$ for particles colliding with $p$. Similarly, $p$ in the fourth quadrant (lower right) is deflected into $p'$ the first quadratic. The signs of $\d C$ the two cases indicate that the collisions carries particles with momentum in the first quadratic into those with momentum in the fourth quadrant.}
		\label{fig_scatterings}
	\end{center}
\end{figure}

Below we argue the displacement current has the same sign as the magnetization current with a simplified collision term. For the steady state reached in a shear flow, the kinetic equation turns into
\begin{align}\label{shear_kinetic}
p^\m\pd_\m f=C[f]\rightarrow p_ip_j\s_{ij}f_V'=\d C,
\end{align}
with $\s_{ij}=\pd_i u_j+\pd_j u_i-\frac{2}{3}\d_{ij}\pd\cdot u$ being the shear tensor.
Without the shear tensor, the collision term vanishes by detailed balance, so we have replaced $C[f]$ by $\d C$, which denotes the collision term induced by the shear.
To be specific, let us consider a shear flow in the $x$-$y$ plane with $\pd_x u_y=\pd_y u_x=\s_{xy}>0$ and restrict our discussion to $p$ in the same plane, see Fig.~\ref{fig_scatterings}. From \eqref{shear_kinetic}, it follows that $\d C<0$ for $p$ in the first and third quadrants ($p_x$ and $p_y$ same sign), while $\d C>0$ for $p$ in the second and fourth quadrants ($p_x$ and $p_y$ opposite sign). The effect of the collision is to bring particles in the quadrants with $\d C<0$ into the quadrants with $\d C>0$.

Without loss of generality, we take $p_x>0$. Thus we have $\d C<0(\d C>0)$ for $p_y>0(p_y<0)$ corresponding to the first and fourth quadrant respectively. The displacement current is given by
\begin{align}\label{disp_S}
S^i_{\text{disp}}\sim\int_{BCD}C_{ABCD}\(\bar{\D}^i_{\l=1/2}-\bar{\D}^i_{\l=-1/2}\)\sim -\d C\e^{ijk}p_jk_k,
\end{align}
where $k$ being average momentum of incoming particles colliding with particle with momentum $p$. To estimate the direction of $k$, we shall make drastic simplification by modeling the collisions by one effective collision with momentum $k$. Let us further consider the limit $p_x\gg p_y$, where the picture is most clear. With the leading logarithmic approximation we consider in this paper, the collision is dominated by small angle scatterings\footnote{The contribution to shear viscosity arises from either multiple small angle scatterings or single large angle scattering \cite{Arnold:2003zc}. The mechanism we discuss applies to small angle scatterings}. $p$ mostly along $x$ in the first quadrant is more probable to be deflected into the fourth quadrant with momentum transfer along negative $y$-direction, as illustrated in Fig.\ref{fig_scatterings}. Similarly $p$ in the third quadrant is more probable to be deflected into the second quadrant with momentum transfer along positive $y$-direction. We will use direction of momentum transfer as the direction of average momentum $k$. Using \eqref{disp_S} and taking into account sign of $\d C$, we obtain displacement current contribution mostly along $z$ direction, with $S^z_{\text{disp}}>0$ for both cases. Meanwhile, we easily obtain the same sign for magnetization current contribution from \eqref{mag}. Analysis for $p$ mostly along $y$ proceeds straightforwardly with both displacement and magnetization contributions flip sign. For $p$ mostly along $x$ but in the fourth quadrant, we have the momentum transfer and $k$ in the positive $y$-direction (also illustrated in Fig.\ref{fig_scatterings} for comparison). In the meantime, $\d C>0$ means that the growth of particles with $p$ in the fourth quadrant is from the first quadrant through collisions. Again in this case, the displacement and magnetization contributions have the same sign. This completes our argument that displacement current contribution enhances that of the magnetization current. Crucially, the argument is based on kinetic equation, which is satisfied only when the particle in question is part of the medium rather than a probe.


\section{Polarization of chiral fermion in QED medium with shear flow}\label{sec_qed}

\subsection{Conducting current contribution}
The conducting current contribution is given by
\begin{align}
p^i(f_R-f_L)=2p^i f_A.
\end{align}
$f_A$ vanishes in equilibrium state. With the state perturbed by shear flow, $f_A$ can be nonvanishing as response to the shear. As in the previous section, we choose $n^\m=u^\m$ and work in the rest frame of the fluid. In this case, it is not possible to construct pseudoscalar $f_A$ out of $\s_{ij}$ and $p_{i}$. It follows that $f_A=0$\footnote{On the contrary, nonvanishing $f_A$ can be generated in a flow with vorticity with $f_A\sim p_i\o_i$.}. The vanishing of spin contribution is due to the interplay of the massless limit and the frame choice $n^\m=u^\m$. Relaxing either of the conditions changes the conclusion. For massive particle, spin becomes an independent degree of freedom \cite{Hattori:2019ahi,Weickgenannt:2019dks,Gao:2019znl}. The massive counterpart of the conducting current becomes a dynamical contribution, which needs to be determined by solving collisional axial kinetic equation, see \cite{Fang:2023bbw,Fang:2022ttm,Wang:2022yli} for recent attempts. For frames with $n^\m\ne u^\m$, we can have $f_A\sim \e^{ijk}p_in_j\s_{kl}p_l$.

\subsection{Displacement and magnetization current contributions}
We move on to evaluate the remaining contributions for a steady state in the shear flow. In order to treat collision effect systematically, we use the collisional quantum kinetic theory \cite{Yang:2020hri,Lin:2021mvw,Hattori:2019ahi,Sheng:2021kfc,Weickgenannt:2021cuo}. The key quantity is the Wigner function defined as \footnote{One could have used Wigner function for Weyl fermions. Since the collision term still involve both chiralities, we choose to work with Dirac fermions and take the massless limit.}
\begin{align}
&W_{\a\b}(X,P)=-\int d^s e^{iP\cdot s}\<\bar{\ps}_{\b}(X-\frac{s}{2})\ps_{\a}(X+\frac{s}{2})\>.
\end{align}
The spin polarization is given by the axial vector component of the lesser Wigner function
\begin{align}
{\cal A}^\m(X,P)=-\frac{1}{4}\tr \big[\g^\m\g^5W(X,P)\big],
\end{align}
whose non-dynamical contribution reads 
\begin{equation}
\label{Amu}
\mathcal{A}^{\mu}=2\pi\delta(P^2)S^{\mu\nu}\mathcal{D}_\nu f,
\end{equation}
where $\mathcal{D}_\nu f=\partial_\nu f-\Sigma_{\nu}^>f-\Sigma_{\nu}^<(1-f)$, and $\Sigma_{\nu}^\gtrless$ are the greater/lesser vector component of fermion self-energy defined as $\S^{\gtrless}_\n=\frac{1}{4}\tr\big[\g_\n\S^{\gtrless}\big]$. We can identify respectively the partial derivative term and self-energy terms as the magnetization and displacement current contributions, i.e. the second and third terms on the RHS of \eqref{current} \cite{Lin:2022tma}. \eqref{Amu} can be viewed as a generalization of those terms in \eqref{current}, which was derived for a system consisting of one species of Weyl fermions.

As we remarked before, $f_A$ vanishes in our case, so $f=f_R=f_L$. For the magnetization contribution, we may simply use the local equilibrium distribution function. In contrast, the displacement contribution needs to be evaluated in a steady state driven by the shear flow. The deviation of the distribution function from the local equilibrium one is crucial for nonvanishing collision term, which is responsible for viscous damping of shear wave \cite{Arnold:2000dr,Arnold:2002zm,Arnold:2003zc}. The same deviation is responsible for displacement contribution \cite{Lin:2022tma}. As has been shown in \cite{Lin:2021mvw}, when spin is not an independent degree of freedom, $f$ satisfies the same kinetic equation as the spin-averaged kinetic theory \cite{Arnold:2000dr,Arnold:2002zm,Arnold:2003zc}.

For illustration purpose, we consider a neutral QED plasma with $N_f$ flavor of massless fermions in a shear flow. We shall further use leading logarithmic (LL) approximation, in which only elastic scatterings among fermions and photons are relevant. The corresponding distribution function has been solved in \cite{Arnold:2000dr,Arnold:2002zm} and approximate analytic expressions have been obtained in \cite{Lin:2022tma}. Below we simply quote the results. Denoting $f$ and $f_\g$ as distribution functions for fermions and photons respectively, we can parameterize the deviations as
\begin{eqnarray}
\label{offeq}
&f_p=f^{\text{eq}}_p+\d f_p,\qquad {f}_{\g,p}={f}^{\text{eq}}_{\g,p}+\d {f}_{\g,p},\no
&\delta f_p=\b f^{\text{eq}}_p\(1-f^{\text{eq}}_p\)I^p_{ij}\sigma_{ij}\chi_p,\qquad \delta {f}_{\g,p}=\b{f}^{\text{eq}}_{\g,p}\(1+{f}^{\text{eq}}_{\g,p}\)I^p_{ij}\sigma_{ij}\gamma_p,
\end{eqnarray}
with $\d f_p$ and $\d{f}_{\g,p}$ being the deviation from local equilibrium distributions $f^\text{eq}_p$ and ${f}^{\text{eq}}_{\g,p}$, which reduce to Fermi-Dirac and Bose-Einstein distributions respectively in local rest frame of fluid. $I^p_{ij}=\hat{p}_i\hat{p}_j-\frac{1}{3}\delta_{ij}$ is a symmetric traceless tensor defined with 3-momentum $p$. 
For $p\gg T$, we have the following approximate solution \cite{Lin:2022tma}
\begin{align}\label{approx_dev}
&\chi_p=\frac{(2\pi)^3}{e^4\ln e^{-1}}\beta^2p^2C_f,\no
&\gamma_p=\frac{(2\pi)^3}{e^4\ln e^{-1}}\beta^2p^2\(\frac{1+{f}_{\g,p}}{1-{f}_p}(C_b-C_f)+C_f\),
\end{align}
where $C_f=\frac{3(1+2N_f)}{4\pi^2N_f^2}$ and $(C_b-C_f)=\frac{2}{\pi^2N_f}$.
In the large $N_f$ limit, we have $C_f\simeq\frac{3}{2\pi^2N_f}$ and $C_b\simeq\frac{7}{2\pi^2N_f}$.
\begin{figure}[htbp]
  \begin{center}
      \includegraphics[height=4.2cm,clip]{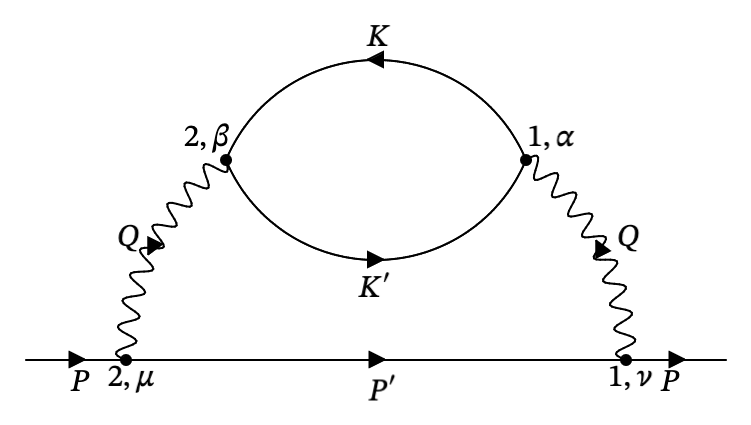}
      \includegraphics[height=4.2cm,clip]{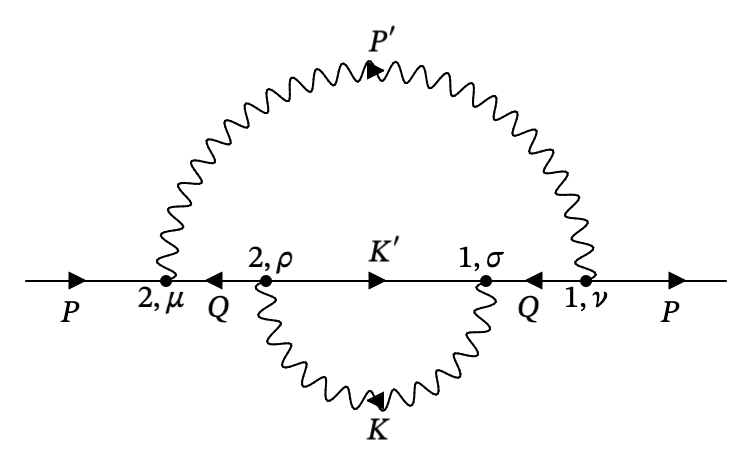}
      \caption{Two-loop diagrams for fermion self-energy, with arrows indicating momentum flow and $1,\;2$ labeling fields on Schwinger-Keldysh contour. In the LL approximation, diagrams with vertex correction are irrelevant. The left diagram gives rise to Coulomb scattering, and the right diagram gives rise to Compton scattering and pair annihilation.}
    \label{fig1}
  \end{center}
\end{figure}

Now we are ready to evaluate the displacement contribution, which appears through self-energy correction $-\Sigma_{\nu}^>f-\Sigma_{\nu}^<(1-f)$ in \eqref{Amu}. We shall consider spin polarization for a fermion with momentum $P$ and take $p_0>0$. To the LL order, the self-energy correction includes Coulomb, Compton and annihilation process which are depicted in Fig.\ref{fig1}. We evaluate self-energy diagram in Fig.~\ref{fig1} as 
\begin{eqnarray}\label{Sigma}
\Sigma_{\text{left}}^>(P)&=&e^4N_f\int_{K,K',P'}(2\pi)^4\delta^4(P+K-P'-K')\nonumber\\
&&\times\gamma^\mu S^>(P')\gamma^\nu D_{\mu\beta}^{22}(-Q)D_{\alpha\nu}^{11}(-Q)\text{tr}\big[\gamma^\alpha S^<(K)\gamma^\beta S^>(K')\big], \\
\Sigma_{\text{right}}^>(P)&=&e^4\int_{K,K',P'}(2\pi)^4\delta^4(P+K-P'-K')\nonumber\\
&&\times \gamma^\mu S_{22}(Q)\gamma^\rho S^>(K')\gamma^\sigma S_{11}(Q)\gamma^\nu D^<_{\rho\sigma}(K)D^>_{\mu\nu}(P'),
\end{eqnarray}
with $\int_P=\int\frac{d^4P}{(2\pi)^4}$ and $Q=P'-P$. $N_f$ in $\Sigma_{\text{left}}^>(P)$ comes from $N_f$ flavors in the fermion loop in the left diagram of Fig.~\ref{fig1}. The real time fermion propagators appearing in \eqref{Sigma} are given by  
\begin{eqnarray}
S^>(P)&=&2\pi\epsilon(p_0){\slashed P} (1-f(p_0)) \delta(P^2),\nonumber\\
S^<(P)&=&2\pi\epsilon(p_0){\slashed P} (-f(p_0)) \delta(P^2),\nonumber\\
S_{22}(P)&=&\;\;\, iS_A(P)+S^<(P),\nonumber\\
S_{11}(P)&=&-iS_R(P)+S^<(P).
\end{eqnarray}
We have splitted $S_{11/22}$ into an on-shell part $S^<(P)$ and off-shell part $S_{A/R}(P)$. The on-shell part is irrelevant because $S_{11/22}$ enters only in Compton scattering and pair annihilation, in which the exchanged fermion is off-shell. We need the following retarded and advanced propagators
\begin{align}
S_R(P)=\frac{-1{\slashed P}}{P^2+i\h\,\e(p_0)},\qquad S_A(P)=\frac{-1{\slashed P}}{P^2-i\h\,\e(p_0)}.
\end{align}
The $i\h$ prescription is irrelevant by the off-shell condition.
We use the Feynman gauge for the real time photon propagators in \eqref{Sigma}
\begin{eqnarray}
&&D_{\mu\nu}^>(P)=2\pi\epsilon(P\cdot u)\delta(P^2)(-g_{\mu\nu})(1+f_{\gamma,p}),\nonumber\\
&&D_{\mu\nu}^<(P)=2\pi\epsilon(P\cdot u)\delta(P^2)(-g_{\mu\nu})f_{\gamma,p},\nonumber\\
&&D_{\mu\nu}^{22}=\frac{ig_{\mu\nu}}{Q^2-i\h},\qquad D_{\mu\nu}^{11}=\frac{-ig_{\mu\nu}}{Q^2+i\h}.
\end{eqnarray}

The vector component of self-energy is obtained through
$\Sigma^{>\lambda}=\frac{1}{4}\text{tr}[\Sigma^> (P)\gamma^\lambda]$. For $\Sigma_{\text{left}}$, we need to evaluate the following traces
\begin{eqnarray}
\text{tr}\big[\gamma^\mu S^>(P')\gamma^\nu\gamma^\lambda]&=&
4(P'^{\mu}g^{\nu\lambda}-P'^{\lambda}g^{\mu\nu}+P'^{\nu}g^{\mu\lambda})2\pi\epsilon(p'_0)\delta(P'^2)(1-f_{p'}), \nonumber\\
\text{tr}\big[\gamma^\alpha S^<(K)\gamma^\beta S^>(K')\big]&=&
4(K^{\alpha} K'^{\beta}+K^{\beta} K'^{\alpha} -K\cdot K' g^{\alpha\beta})\nonumber\\
&&\times (2\pi)^2\epsilon(k_0)\epsilon(k'_0) (-f_k)(1-f_{k'}) \delta(K^2) \delta(K'^2).
\end{eqnarray}
In the LL approximation $Q\ll K,\,P$, so that $\epsilon(k_0)\epsilon(k'_0)\simeq 1$ and $\e(p_0')\simeq1$. The cases with $k_0$ and $k_0'$ both positive and negative correspond to Coulomb scatterings with fermion and anti-fermions respectively, which give identical contributions in a charge neutral plasma.
Thus the self-energy corresponding to Coulomb scattering gives the vector component
\begin{align}\label{coul}
\Sigma^{>}_{Coul,\mu}(P)
=&-4e^4(2\pi)^3\int_{K,K',P'}(2\pi)^4\delta^4(P+K-P'-K') \nonumber\\
&\times\epsilon(p'_0)\epsilon(k_0)\epsilon(k'_0)\delta(P'^2)\delta(K^2) \delta(K'^2)\nonumber\\
&\times\frac{2N_f\cdot2 [K_\mu P'\cdot K'+K'_\mu P'\cdot K]}{(Q^2)^2}
f_k(1-f_{k'})(1-f_{p'}),
\end{align}
the factor $2N_f$ corresponding to $N_f$ flavor of fermions and anti-fermion in the loop.
For $\Sigma_{\text{right}}$, we need to evaluate the following traces
\begin{eqnarray}
&&\text{tr}\big[\gamma^\mu S_{22}(Q)\gamma^\rho S^>(K')\gamma^\sigma S_{11}(Q)\gamma^\nu D^<_{\rho\sigma}(K)D^>_{\mu\nu}(P')\gamma^\lambda\big]\nonumber\\
&=&16\big(2K'\cdot QQ^\lambda-K'^\lambda Q^2\big)\frac{1}{(Q^2)^2}\nonumber\\
&&\times(2\pi)^3\epsilon(k'_0)\epsilon(k_0)\epsilon(p'_0)\delta(P'^2)\delta(K^2) \delta(K'^2)f_{\gamma,k}(1-f_{k'})(-1-f_{\gamma,p'}).
\end{eqnarray}
Similar to the analysis of $\Sigma_{\text{left}}$, we have $\epsilon(k_0)\epsilon(k'_0)\simeq 1$ and $\e(p_0')\simeq1$. But now the case with $k_0$ and $k_0'$ both positive and negative correspond to Compton scatterings and pair annihilation respectively.
The self-energy corresponding to Compton scattering gives 
\begin{align}\label{comp}
\Sigma^{>}_{Comp,\mu}(P)
=&-4e^4(2\pi)^3\int_{K,K',P'}(2\pi)^4\delta^4(P+K-P'-K') \nonumber\\
&\times\epsilon(p'_0)\epsilon(k_0)\epsilon(k'_0)\delta(P'^2)\delta(K^2) \delta(K'^2)\nonumber\\
&\times\frac{[2K'\cdot QQ_\mu-K'_\mu Q^2]}{(Q^2)^2}
f_{\gamma,k}(1-f_{k'})(1+f_{\gamma,p'})
\end{align}
Annihilation process can be obtained through implementing $K\rightarrow -K$ as well as $K'\rightarrow -K'$ in the above expression.

The combination needed for polarization is $-f_p\Sigma^>_\mu(P)-(1-f_p)\Sigma^<_\mu (P)$. 
We have shown above that $\Sigma^>_\mu(P)$ contains contribution from Coulomb scattering, Compton scattering and annihilation at LL order, namely $\Sigma^>_\mu=\Sigma^{>}_{Coul,\mu}+\Sigma^{>}_{Comp,\mu}+\Sigma^{>}_{anni,\mu}$. With \eqref{coul} and \eqref{comp}, we easily identify the term $-f_p\Sigma^>_\mu(P)$ from the product of distribution functions in as loss term. Similar analysis allows us to identify the term $-(1-f_p)\Sigma^<_\mu (P)$ as gain term. Crucially they do not cancel in the steady state, which can be seen in the following explicit expression
\begin{align}\label{fSigma}
&-f_p\Sigma^{>}_{\mu}(P)-(1-f_p)\Sigma^{<}_{\mu}(P)\nonumber\\
=\;\;&4e^4(2\pi)^3\int_{K,K',P'}(2\pi)^4\delta^4(P+K-P'-K') \delta(P'^2)\delta(K^2) \delta(K'^2)\nonumber\\
&\times\Big[M_\mu^{Coul}
\big(I_{ij}(\hat{p})\chi_p+I_{ij}(\hat{k})\chi_k-I_{ij}(\hat{k'})\chi_{k'}-I_{ij}(\hat{p'})\chi_{p'}\big)f^\eq_pf^\eq_k(1-f^\eq_{k'})(1-f^\eq_{p'})\nonumber\\
&+M_\mu^{Comp}
\big(I_{ij}(\hat{p})\chi_p+I_{ij}(\hat{k})\gamma_k-I_{ij}(\hat{k'})\chi_{k'}-I_{ij}(\hat{p'})\gamma_{p'}\big)f^\eq_pf^\eq_{\gamma,k}(1-f^\eq_{k'})(1+f^\eq_{\gamma,p'})\nonumber\\ 
&+ M_\mu^{anni}
\big(I_{ij}(\hat{p})\chi_p+I_{ij}(\hat{k'})\chi_{k'}-I_{ij}(\hat{k})\gamma_k-I_{ij}(\hat{p'})\gamma_{p'}\big)f^\eq_pf^\eq_{k'}(1+f^\eq_{\gamma,k})(1+f^\eq_{\gamma,p'})\Big]\sigma^{ij},
\end{align}
where we have linearized in the shear tensor and defined the following quantities
\begin{align}\label{amplitude}
M_\mu^{Coul}&=\frac{2N_f \cdot 2[P\cdot K(2K_\mu-Q_\mu)]}{(Q^2)^2},\nonumber\\
M_\mu^{Comp}&=M_\mu^{anni}=-\frac{-K_\mu}{Q^2},
\end{align}
which can be viewed as Coulomb, Compton scattering and pair annihilation amplitude squares times $\bar{\D}_\m$ comparing to \eqref{current}.
To make full use of Lorentz invariance, we rewrite \eqref{fSigma} in fluid's local rest frame to an arbitrary frame. This can be conveniently done with the projector $\Delta^{\mu\nu}=g^{\mu\nu}-u^\mu u^\nu$. The generalizations of $I_{ij}$ and $\s^{ij}$ to an arbitrary frame are given by
\begin{align}
&I_{\alpha\beta}^P=\hat{P}_{\perp\alpha}\hat{P}_{\perp\beta}+\frac{1}{3}\Delta_{\alpha\beta},\nonumber\\
&\s^{\a\b}=\D^{\a\m}\D^{\b\n}\(\pd_\m u_\n+\pd_\n u_\m-\frac{2}{3}\D_{\m\n}\pd_l u^\l\),
\end{align}
with $P_\perp^\mu=\Delta^{\mu\nu}P_\nu$ and $\hat{P}_\perp^\mu=P_\perp^\mu/p$. Furthermore, we may replace everywhere $p\to (-P_\perp^\mu P_{\perp\mu})^{1/2}$ and similarly for other momenta in \eqref{fSigma}. Since the distribution functions in spin-averaged kinetic theory are Lorentz scalars, we deduce that \eqref{fSigma} is a Lorentz vector, which can be parameterized as
\begin{align}
-f_p\Sigma^{>}_{\mu}(P)-(1-f_p)\Sigma^{<}_{\mu}(P)=\;T_{\mu,\alpha\beta}\sigma^{\alpha\beta}
\end{align}
The coefficient $T_{\mu,\alpha\beta}$ can be chosen to share the same property as $\s^{\a\b}$ in the last two indices: symmetric, traceless and orthogonal to fluid velocity. It can be decomposed into the following structures
\begin{eqnarray}
\label{self-energy-decompose}
T_{\mu,\alpha\beta}&=&(u_\mu I_{\alpha\beta}T_{1}+\hat{P}_{\perp\mu}I_{\alpha\beta}T_{2}+J_{\mu,\alpha\beta}T_{3})f^\eq_p(1-f^\eq_p),
\end{eqnarray}
with the last structure defined as
\begin{eqnarray}
\label{projector1}
J_{\mu,\alpha\beta}^P&=&-\hat{P}_{\perp\alpha}\Delta_{\mu\beta}-\hat{P}_{\perp\beta}\Delta_{\mu\alpha}+\frac{2}{3}\hat{P}_{\perp\mu}\Delta_{\alpha\beta}.
\end{eqnarray}
The factor $f^\eq_p(1-f^\eq_p)$ is introduced for convenience.

In \eqref{Amu}, 
$-f_p\Sigma^{>}_{\mu}-(1-f_p)\Sigma^{<}_{\mu}$ is contracted with $\epsilon_{\mu\nu\rho\sigma}P^\rho u^\sigma$, hence only $J_{\mu,\alpha\beta}T_{3}$ in \eqref{self-energy-decompose} contributes to the polarization, while $u_\mu I_{\alpha\beta}$ and $\hat{P}_{\perp\mu}I_{\alpha\beta}$ vanish upon contraction. We then obtain the following displacement contribution to polarization
\begin{align}\label{A_disp}
\mathcal{A}_\mu^\Sigma&=2\pi \delta(P^2) S_{\mu\nu}(-f_p\Sigma^{>\nu}-\bar{f}_p\Sigma^{<\nu})\nonumber\\
&=-2\pi \delta(P^2)\frac{\epsilon_{\mu\nu\alpha\beta}P_\perp^\alpha u^\beta}{2P\cdot u}P_{\perp\xi}\sigma^{\nu\xi}\frac{2T_{3}}{p}f^\eq_p(1-f^\eq_p).
\end{align}
This is to be compared with the magnetization contribution implemented in phenomenological studies \cite{Hidaka:2017auj,Liu:2021uhn,Becattini:2021suc}.
\begin{align}\label{A_magn}
\mathcal{A}_\mu^\partial=2\pi \delta(P^2) S_{\mu\nu}\partial^\nu f_p=-2\pi \delta(P^2)\frac{\epsilon_{\mu\nu\alpha\beta}P_\perp^\alpha u^\beta}{2P\cdot u}P_{\perp\xi}\sigma^{\nu\xi}f^\eq_p(1-f^\eq_p).
\end{align}
%
$T_{3}$ can be obtained by projecting out the corresponding component in $T_{\mu,\alpha\beta}$ by suitable combination of projectors and then performing the momentum integrals. We collect details in Appendix A. Using \eqref{approx_dev}, we obtain the following analytic expression
\begin{eqnarray}\label{T3}
T_3&=&\frac{C_f N_f \left(9 T \zeta (3) (4 T-5 F p)+\pi ^2 p (F p-4 T)\right)}{6 p}\nonumber\\
&&+\frac{(C_b-C_f)\pi ^4 T^2 }{48 p} \frac{1+f^\eq_{\gamma, p}}{1-f^\eq_p},
\end{eqnarray}
with $F=1-2f^\eq_p$. We observe that $T_3$ splits into contribution from Coulomb scattering (first line) and from Compton and annihilation (second line).

The ratio $R^{\Sigma/\partial}\equiv\frac{2T_{3}}{p}$ delineates the importance of displacement current which appears as $A_\mu^{\Sigma}$ in comparison to the well investigated magnetization current contribution $A_\mu^{\partial}$. It is instructive to look at contribution from Coulomb and Compton/annihilation separately. The sign of the Coulomb contribution depends on $p$. In the limit $p\ll T$, we have from the Coulomb contribution
\begin{align}
R_{\text{Coul}}^{\Sigma/\partial}(p\ll T)\simeq \frac{3C_f N_f T^2\zeta(3)}{p^2}.
\end{align}
The positive sign corresponds to an enhancement effect. Note that the regime of soft $p$ is where we expect the single collision approximation used in Sec.~\ref{sec_enhance} to become accurate and indeed our explicit calculations gives an enhancement effect consistent with our argument in Sec.~\ref{sec_enhance}. Physical interpretation is that softer particle is more easily polarized through collisional effect. The apparent infrared divergence will not occur as very soft particles lie outside the domain of kinetic theories.
In the phenomenologically interesting limit $p\gg T$, we have from the Coulomb contribution
\begin{align}
R_{\text{Coul}}^{\Sigma/\partial}(p\gg T)\simeq \frac{C_f N_f \pi ^2}{3}.
\end{align}
It also leads to an enhancement of the magnetization contribution. For not very large $p$, the ratio can be negative corresponding to suppression of the magnetization contribution. Interestingly, the sign of the Compton/annihilation contribution is always positive. We do not have naive generalization of \eqref{current} for these processes, which requires inclusion of gauge particles to the kinetic theory.

\begin{figure}[htbp]
\begin{center}
\includegraphics[height=5cm,clip]{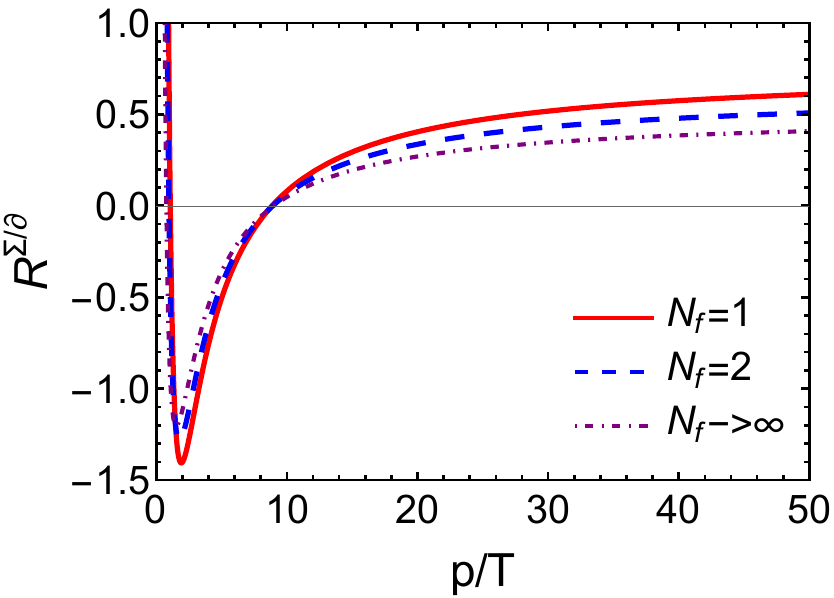}
\caption{The ratio $R^{\Sigma/\partial}\equiv\frac{2T_{3}}{p}$ of the contributions from the displacement current and the magnetization current. In the limits $p\ll T$ and $p\gg T$, the displacement current contribution enhances the magnetization contribution. For intermediate $p$, the displacement current contribution leads to suppression of the magnetization current contribution.}
\label{fig2}
\end{center}
\end{figure}

In Fig.\ref{fig2}, the ratio $R^{\Sigma/\partial}\equiv\frac{2T_{3}}{p}$ is plotted as a function of $p/T$, for one, two flavor as well as $N_f\rightarrow \infty$. In the phenomenologically interesting limit $p\gg T$, we have $R^{\Sigma/\partial}\simeq R_{\text{Coul}}^{\Sigma/\partial}=\frac{1}{2}\Big(1+\frac{1}{2N_f}\Big)$, using explicit expression of $C_f$ below \eqref{approx_dev}. It indicates that the contribution from displacement current is comparable to the counterpart from magnetization current.

\section{Conclusion and Outlook}\label{sec_conclude}

We have considered spin polarization of chiral fermion in a shear flow. We have made close connection with the collisional chiral kinetic theory, identifying conducting, magnetization and displacement current contributions to spin polarization. While each contribution is frame dependent, their sum is not. In the frame set by the fluid's rest frame, we find the conducting current contribution vanishes. The magnetization current contribution widely used in phenomenological studies is independent of interaction. The displacement current contribution does depends on interaction. It is nonvanishing in the steady state reached in the shear flow. We determine this contribution with QED interaction, in which we have found nontrivial momentum dependence: the displacement current contribution enhances the magnetization current contribution for fermions at small and large momenta, but leads to a suppression effect at intermediate momenta. This might help understand the local polarization measurement of $\L$ hyperon in heavy ion collisions.

The massless limit plays a special role in our study. In this limit, the spin is enslaved by fermion momentum, leading to the vanishing of spin contribution to polarization in a shear flow. This will not be true for massive fermions. An obvious generalization is to consider massive fermions. A qualitative new feature is that the massive counterpart of the conducting current contribution needs to be determined dynamically. In the steady state reached in a shear flow, one can still determine it by solving the axial kinetic equation. We leave this for future studies.

\section*{Acknowledgments}

We are grateful to Shi Pu and Di-Lun Yang for insightful discussions and comments on an earlier version of the paper. This work is in part supported by NSFC under Grant Nos 12075328, 11735007 (S.L.) and 12005112 (Zy.W.).

\appendix

\section{Evaluation of displacement contribution to polarization}
In this section, we collect the key procedure in obtaining $T_3$ which determines the displacement contribution to polarization as defined in \eqref{A_disp}. From the decomposition \eqref{self-energy-decompose}, $T_3$ can be obtained from projecting $T_{\mu,\alpha\beta}$ to combinations of projectors 
\begin{align}
T_3=\frac{1}{4f_p(1-f_p)}(2\hat{P}_{\perp}^\mu I^{\alpha\beta}-J^{\mu,\alpha\beta})T_{\mu,\alpha\beta}
\end{align}
with $T_{\mu,\alpha\beta}$ presented in \eqref{fSigma} contains contributions from Coulomb scattering, Compton scattering and annihilation. We need to evaluate the following contractions for Coulomb scattering 
\begin{align}
I^{Coul}
=&(2\hat{P}_{\perp}^\mu I^{\alpha\beta}-J^{\mu,\alpha\beta})M_\mu^{Coul}
\big(I_{\alpha\beta}(\hat{p})\chi_p+I_{\alpha\beta}(\hat{k})\chi_k-I_{\alpha\beta}(\hat{k'})\chi_{k'}-I_{\alpha\beta}(\hat{p'})\chi_{p'}\big)\nonumber\\
&\times f^\eq_pf^\eq_{k'}(1+f^\eq_{\gamma,k})(1+f^\eq_{\gamma,p'})\nonumber\\
=&\frac{8N_f P\cdot K}{(Q^2)^2}\Big\{\Big[2k\,\hat{p}\cdot \hat{k}\Big(1-(\hat{p}\cdot \hat{k})^2\Big)
-q\Big(\hat{p}\cdot \hat{k}\hat{k}\cdot \hat{q}-(\hat{p}\cdot \hat{k})^2\hat{p}\cdot \hat{q}\Big)\Big]\chi_k
\nonumber\\
&
-\Big[2k\Big(\hat{p}\cdot \hat{k}'\hat{k}\cdot \hat{k}'-\hat{p}\cdot \hat{k}(\hat{p}\cdot \hat{k}')^2\Big)
-q\Big(\hat{p}\cdot \hat{k}'\hat{k}'\cdot \hat{q}-(\hat{p}\cdot \hat{k}')^2\hat{p}\cdot \hat{q}\Big)\Big]\chi_{k'}
\nonumber\\
&
-\Big[2k\Big(\hat{p}\cdot \hat{p}'\hat{k}\cdot \hat{p}'-\hat{p}\cdot \hat{k}(\hat{p}\cdot \hat{p}')^2\Big)
-q\Big(\hat{p}\cdot \hat{p}'\hat{p}'\cdot \hat{q}-(\hat{p}\cdot \hat{p}')^2\hat{p}\cdot \hat{q}\Big)\Big]\chi_{p'}\Big\}\nonumber\\
&\times f^\eq_pf^\eq_{k'}(1+f^\eq_{\gamma,k})(1+f^\eq_{\gamma,p'}),
\end{align}
with $M_\mu^{Coul}$ defined in \eqref{amplitude}. For Compton scattering and for annihilation, the following contractions are required
\begin{align}
I^{Comp}=&(2\hat{P}_{\perp}^\mu I^{\alpha\beta}-J^{\mu,\alpha\beta})M_\mu^{Comp}
\big(I_{\alpha\beta}(\hat{p})\chi_p+I_{\alpha\beta}(\hat{k})\gamma_k-I_{\alpha\beta}(\hat{k'})\chi_{k'}-I_{\alpha\beta}(\hat{p'})\gamma_{p'}\big)\nonumber\\
&\times 
f^\eq_pf^\eq_{\gamma,k}(1-f^\eq_{k'})(1+f^\eq_{\gamma,p'})\nonumber\\
=&2k\Big[+\hat{p}\cdot \hat{k}\Big(1-(\hat{p}\cdot \hat{k})^2\Big)\gamma_k
-\Big(\hat{p}\cdot \hat{k}'\hat{k}\cdot \hat{k}'-\hat{p}\cdot \hat{k}(\hat{p}\cdot \hat{k}')^2\Big)\chi_{k'}\nonumber\\
& 
-\Big(\hat{p}\cdot \hat{p}'\hat{k}\cdot \hat{p}'-\hat{p}\cdot \hat{k}(\hat{p}\cdot \hat{p}')^2\Big)\gamma_{p'}\Big]f^\eq_pf^\eq_{\gamma,k}(1-f^\eq_{k'})(1+f^\eq_{\gamma,p'}),\\
I^{anni}=&(2\hat{P}_{\perp}^\mu I^{\alpha\beta}-J^{\mu,\alpha\beta})M_\mu^{anni}
\big(I_{\alpha\beta}(\hat{p})\chi_p+I_{\alpha\beta}(\hat{k'})\chi_{k'}-I_{\alpha\beta}(\hat{k})\gamma_k-I_{\alpha\beta}(\hat{p'})\gamma_{p'}\big)\nonumber\\
&\times f^\eq_pf^\eq_{k'}(1+f^\eq_{\gamma,k})(1+f^\eq_{\gamma,p'})\nonumber\\
=&2k\Big[
-\hat{p}\cdot \hat{k}\Big(1-(\hat{p}\cdot \hat{k})^2\Big)\gamma_k
+\Big(\hat{p}\cdot \hat{k}'\hat{k}\cdot \hat{k}'-\hat{p}\cdot \hat{k}(\hat{p}\cdot \hat{k}')^2\Big)\chi_{k'}
\nonumber\\
&
-\Big(\hat{p}\cdot \hat{p}'\hat{k}\cdot \hat{p}'-\hat{p}\cdot \hat{k}(\hat{p}\cdot \hat{p}')^2\Big)\gamma_{p'}\Big]f^\eq_pf^\eq_{k'}(1+f^\eq_{\gamma,k})(1+f^\eq_{\gamma,p'})
\end{align}
with $M_\mu^{Comp}$ and $M_\mu^{anni}$ defined in \eqref{amplitude}.

On the other hand the phase space integral in \eqref{fSigma} can be simplified under the condition of  small momentum transfer
\begin{align}
\label{measure1}
&4e^4(2\pi)^3\int_{K,K',P'}(2\pi)^4\delta^4(P+K-P'-K') \delta(P'^2)\delta(K^2) \delta(K'^2)\nonumber\\
=&4e^4\int\frac{d^4Kd^4K'd^4P'd^4Q}{(2\pi)^{5}}\delta(P+Q-P')\delta(K'+Q-K)\delta(P'^2)\delta(K^2) \delta(K'^2)\nonumber\\
=&\frac{4e^4}{(2\pi)^5}\int dq_0 d^3qd^3k\frac{1}{2p'_02k'_02k_0}\delta(p_0-p'_0+q_0)\delta(k_0-k'_0-q_0).
\end{align}
It is useful to decompose momentum $\vec{q}$ and $\vec{k}$ into
\begin{align}
\label{momentum_decomposition}
\vec{q}&= q\cos\theta\hat{p}+q\sin\theta\cos\varphi\hat{x}+q\sin\theta\sin\varphi\hat{y},\nonumber\\
\vec{k}&= k\cos\theta'\hat{p}+k\sin\theta'\cos\varphi'\hat{x}+k\sin\theta'\sin\varphi'\hat{y},
\end{align}
$\hat{p}$ is chosen as the $\hat{z}$-direction. $\Omega$ is introduced as the angle between $\vec k$ and $\vec q$, namely $\cos\Omega=\cos\theta'\cos\theta-\sin\theta'\sin\theta\cos\Delta\varphi$, with $\Delta\varphi=\varphi-\varphi'$.  The measure can be parameterized as 
\begin{eqnarray}
\int d^3qd^3k=\int q^2 dq d\cos\theta d\varphi k^2 dk d\cos\theta' d\varphi'.
\end{eqnarray}
Using the decomposition above as well as the on-shell condition, to $\mathcal{O}(q^2)$, the $\delta$-functions in the integral can be casted into 
\begin{align}
\delta(p_0-p'_0+q_0)&\simeq\delta(-q\cos\theta-\frac{q^2}{2p}\sin^2\theta+q_0),\nonumber\\
\delta(k_0-k'_0-q_0)&\simeq\delta(q\cos\Omega-\frac{q^2}{2k}\sin^2\Omega-q_0).
\end{align}
The angular integral over $\varphi$ and $\varphi'$ can be performed with the help of the above $\delta$-functions
\begin{align}
\int d\varphi d\varphi'\delta(k_0-k'_0-q_0)&\simeq&\frac{4\pi}{q(1+\frac{q_0}{k})}\frac{1}{[\sin^2\theta\sin^2\theta'-(\cos\Omega-\cos\theta\cos\theta')^2]^{1/2}}.
\end{align}
The square root constrains the domain of $\cos\theta'$ as $\cos(\theta-\Omega)<\cos\theta'<\cos(\theta+\Omega)$. The other $\delta$-function gives 
\begin{align}
\int d\cos\theta\delta(p_0-p'_0+q_0)\simeq\frac{1}{q(1-\frac{q_0}{p})}.
\end{align}
The $\delta$-functions also fix the angles to $\mathcal{O}(q^2)$ as 
\begin{align}
&\cos\Omega=\frac{q_0}{q}+\frac{q}{2k}\Big(1-\frac{q_0^2}{q^2}\Big),\quad
\sin\Omega=\Big(1-\frac{q_0^2}{q^2}\Big)^{1/2}\Big(1-\frac{q_0}{2{k}}\Big),\nonumber\\
&\cos\theta=\frac{q_0}{q}-\frac{q}{2p}\Big(1-\frac{q_0^2}{q^2}\Big), \quad\;\;
\sin\theta=\Big(1-\frac{q_0^2}{q^2}\Big)^{1/2}\Big(1+\frac{q_0}{2p}\Big).
\end{align}
With all the pieces above, the coefficient $T_3$ can be obtained through completing the following momentum integral 
\begin{align}
T_3
=\frac{e^4}{4(2\pi)^4p}\int\frac{dq_0 dq dk d\cos\theta'}{[\sin^2\theta\sin^2\theta'-(\cos\Omega-\cos\theta\cos\theta')^2]^{1/2}}
\frac{(I^{Coul}+I^{Comp}+ I^{anni})}{f^\eq_p(1-f^\eq_p)}.
\end{align}
To proceed, we expand the integrand to $O(q^{-2})$. The resulting $\th'$ and $k$ integrals can be performed analytically. To arrive at the leading logarithmic result, we integrate $q_0$ over $(-q,q)$ and then $q$ in the range $eT\ll q\ll T$.
The logarithm thus arises from $\int_{eT}^{T}dq/q=\ln(1/e)$. The overall factor $e^4\ln(1/e)$ then cancels out those in $\chi_p$ and $\gamma_p$ in the deviation from local equilibrium distributions \eqref{offeq}, yielding the result of $T_3$ in \eqref{T3} independent of coupling constant.

\bibliographystyle{unsrt}
\bibliography{shear_massless.bib}

\end{document}